\documentclass[12pt]{article}
\usepackage{graphicx}
\usepackage{amssymb}
\usepackage{amsmath}
\usepackage{bm}

\newcommand{\bear}{\begin{array}}  
\newcommand {\eear}{\end{array}}
\newcommand{\bea}{\begin{eqnarray}}   
\newcommand{\eea}{\end{eqnarray}}
\newcommand{\beq}{\begin{eqnarray}}   
\newcommand{\eeq}{\end{eqnarray}}
\newcommand{\bef}{\begin{figure}}  \newcommand 
{\eef}{\end{figure}}
\newcommand{\bec}{\begin{center}}  \newcommand 
{\eec}{\end{center}}

\newcommand{\Slash}[1]{{\ooalign{\hfil/\hfil\crcr$#1$}}}

\begin{document}

\begin{titlepage}

\begin{flushright}
IPMU12-0074 
\end{flushright}

\vskip 1.35cm
\begin{center}

  {\large {\bf  Enhancement of Proton Decay Rates \\in
      Supersymmetric SU(5) Grand Unified Models
 }
}

\vskip 1.2cm

Junji Hisano$^{a,b}$, 
Daiki Kobayashi$^a$,
and
Natsumi Nagata$^{a,c}$\\

\vskip 0.4cm
{\it $^a$Department of Physics,
Nagoya University, Nagoya 464-8602, Japan}\\
{\it $^b$IPMU, TODIAS,
University of Tokyo, Kashiwa 277-8568, Japan}\\
{\it $^c$Department of Physics, 
University of Tokyo, Tokyo 113-0033, Japan}

\date{\today}

\begin{abstract} 
  In the supersymmetric grand unified theories (SUSY GUTs), gauge
  bosons associated with the unified gauge group induce proton
  decay. We investigate the proton decay rate via the gauge bosons in
  the SUSY GUTs under the two situations; one is with extra
  vector-like multiplets, and the other is with heavy sfermions. It is
  found that the proton lifetime is significantly reduced in the
  former case, while in the latter case it is slightly
  prolonged. Determination of the particle contents and their mass
  spectrum below the GUT scale is important to predict the proton
  lifetime. The proton decay searches have started to access to the
  $10^{16}$~GeV scale. 
\end{abstract}

\end{center}
\end{titlepage}

\section{Introduction}
\label{introduction}

The grand unified theories (GUTs), which embed the Standard Model (SM)
gauge group into a large single gauge group, are quite attractive, and
so a variety of models of the theories are proposed since the earliest
work based on the SU(5) symmetry group was presented by Georgi and
Glashow in 1974 \cite{Georgi:1974sy}. Among them, the supersymmetric
grand unified theories (SUSY GUTs) are considered to be promising
candidates since they realize the gauge coupling unification with
great accuracy \cite{Dimopoulos:1981yj} as well as solving the
hierarchy problem in the GUTs \cite{Georgi:1974yf,
Dimopoulos:1981zb}. The supersymmetric version of the Georgi-Glashow
SU(5) GUT, which is the simplest among the SUSY GUTs, is called the
Minimal SUSY SU(5) GUT \cite{Dimopoulos:1981zb}.

One of the most distinctive features which GUTs predict is the
existence of baryon-number violating interactions, such as proton
decay. In the SUSY GUTs with $R$ parity, for example, proton decay is
induced by two different processes: the colored-Higgs and the
$X$-boson exchanging processes. The colored Higgs triplets are
introduced for the Higgs doublets in the Minimal Supersymmetric
Standard Model (MSSM) to be incorporated into the SU(5)
multiplets. The colored-Higgs exchange yields baryon-number
violating dimension-five operators, which give rise to the dominant
contribution to proton decay in the Minimal SUSY SU(5) GUTs
\cite{Sakai:1981pk}. On the other hand, $X$ bosons are the gauge
bosons in the unified SU(5) gauge group, and they induce proton
decay through the dimension-six operators. Since there has been no
experimental signal of proton decay so far, strong limits are imposed
on these interactions, especially on the former one. Super-Kamiokande
gives the bounds on the proton lifetime:
$\tau(p\to e^+ \pi^0)> 1.29\times 10^{34}$~yrs  with 219.7 kt-yr of data
\cite{Nishino:2012rv}  and $\tau(p\to K^+ 
\bar{\nu})>3.3\times 10^{33}$~yrs  with 172.8 kt-yr of data at 90\%
confidence level 
\cite{Miura:2010zz}. These bounds are so stringent that the Minimal
SU(5) SUSY GUT is excluded since the colored-Higgs-boson exchange
process yields too short lifetime in the $p\to K^+\bar{\nu}$ channel
\cite{Goto:1998qg}. This prediction is, however, quite
model-dependent. In fact, many models are proposed in order to allow
them to evade the experimental constraints. Such attempts
\cite{Hisano:1992ne} are accomplished by pushing the mass of the
colored Higgs much heavier than the GUT scale and/or suppressing the
dimension-five operators by a certain symmetry, {\it e.g.}, the
Peccei-Quinn (PQ) symmetry \cite{Peccei:1977hh} in the
four-dimensional GUTs.  In the extra-dimensional models, the $R$ symmetry
is introduced to suppress the dimension-five proton decay
\cite{Hall:2001pg}. With the extensions, the $X$-boson exchanging
interactions become dominant and the proton lifetime turns out to be
long enough to avoid the current limit. In this work, we assume such
mechanism works and concentrate on the proton decay via the $X$-boson
exchange. 

Nowadays, the weak-scale supersymmetry itself is severely constrained
by the experiments at the Large Hadron Collider (LHC). Since there has
been no signal of SUSY particles, the ATLAS and the CMS collaborations
give severe limits on their masses, especially those of colored
particles \cite{Aad:2011ib}. Moreover, the collaborations have
recently announced that they detected signatures of Higgs boson for
the first time \cite{:2012si}. According to their results, the mass of
Higgs boson is about 125 GeV. The result is welcome to the MSSM, which
predicts the light Higgs boson, while the substantial radiative
corrections are required in order to raise the mass of Higgs boson up
to 125 GeV. Since the corrections are increased as the SUSY particles
become heavy, this situation implies that the SUSY scale might be
higher than the electroweak scale \cite{Ibe:2011aa}, and challenges
natural solutions to the hierarchy problem. These results give
constraints on models of the SUSY-breaking mediation. The low-energy
gauge mediation model may be required to introduce several messenger
multiplets so that the  SUSY particles are heavy enough
\cite{Ajaib:2012vc}.  
On the other hand, there have been several alternatives to explain the
125 GeV Higgs boson in an extension of the MSSM. Introducing
vector-like supermultiplets to the MSSM is one of the simplest way to
accomplish the purpose \cite{Moroi:1992zk}. In this case, the quantum
effects by the extra multiplets help to increase the Higgs-boson
mass. With the multiplets being a representation of the grand unified
group, the perturbative gauge coupling unification is still preserved
with great accuracy.

The MSSM with vector-like multiplets and high-scale SUSY models
modifies the proton decay rates. In this work, we examine it in
detail. We will find that vector-like matters enhance the proton decay
rate, while heavy sfermions extend its lifetime. This investigation
indicates that the proton-decay experiments might provide a hint for
the low-energy structure of the SUSY GUTs.

We assume that the SUSY GUTs are realized in the four-dimensional
spacetime, and not consider the extra-dimensional SUSY GUTs in which
the gauge symmetry is broken by the boundary conditions. In the latter
models, the main decay modes of proton depend on configuration of
quarks and leptons in the extra-dimensional space
\cite{Hebecker:2002rc}. However, the decrease/enhancement of the proton
lifetime we show are almost universal even in those models.

This paper is organized as follows. We begin by reviewing the SUSY
SU(5) GUTs in Sec.~\ref{msgut}, and discuss the proton decay via the
$X$-boson exchange in the subsequent section. Then, we evaluate the
lifetime of proton decay in the case of the MSSM with vector-like
matters and heavy sfermions in Sec.~\ref{vectorlike} and
Sec.~\ref{heavy_susy}, respectively. Section \ref{conclusion} is
devoted to conclusion. In Appendix, we present formulae for evaluating
the long-distance part of the renormalization factors.

\section{SUSY SU(5) GUTs}
\label{msgut}

First, we review the SUSY SU(5) GUTs
in order to illustrate the notation and conventions which we use in this
article. Just like the Georgi-Glashow SU(5) model \cite{Georgi:1974sy},
we assume that the SM fermions, as well as their superpartners, are
embedded in a $\bar{\bf 5}\oplus {\bf 10}$ representation. The
multiplets $\Phi$ and $\Psi$, which are the matter fields of $\bar{\bf
5}$ and ${\bf 10}$ representations, respectively, are identified as
\begin{align}
 \Phi &=
\begin{pmatrix}
 \bar{D}_1^\prime \\
 \bar{D}_2^\prime \\
 \bar{D}_3^\prime \\
 E^\prime \\
 -N^\prime
\end{pmatrix}~, ~~~~~~
\Psi=\frac{1}{\sqrt{2}}
\begin{pmatrix}
 0&\bar{U}_3^\prime&-\bar{U}_2^\prime&U^{\prime 1}&D^{\prime 1} \\
 -\bar{U}_3^\prime&0&\bar{U}_1^\prime&U^{\prime 2}&D^{\prime 2}\\
 \bar{U}_2^\prime&-\bar{U}_1^\prime&0&U^{\prime 3}&D^{\prime 3}\\
 -U^{\prime 1}&-U^{\prime 2}&-U^{\prime 3}&0&\bar{E}^\prime \\
 -D^{\prime 1}&-D^{\prime 2}&-D^{\prime 3}&-\bar{E}^\prime &0
\end{pmatrix}~,
\end{align}
where all of the component fields are expressed in terms of the
left-handed chiral superfields, and primes show that the above fields
are the gauge eigenstates.
The super/sub scripts of $U^{\prime \alpha}$,
$D^{\prime \alpha}$, $\bar{U}_\alpha^\prime$, and
$\bar{D}_\alpha^\prime$ denote a color index with $\alpha=1,2,3$. The
chiral superfields $E^\prime$ and $N^\prime$ in $\Phi$ as well as
$U^{\prime \alpha}$ and
$D^{\prime \alpha}$ in $\Psi$ form the SU(2)$_L$ doublets, respectively, 
\begin{equation}
 L^\prime=
\begin{pmatrix}
 N^\prime \\ E^\prime 
\end{pmatrix}~, ~~~~~~
Q^{\prime \alpha}=
\begin{pmatrix}
 U^{\prime \alpha} \\ D^{\prime \alpha}
\end{pmatrix}~,
\end{equation}
while $\bar{U}_\alpha^\prime$, $\bar{D}_\alpha^\prime$, and
$\bar{E}^\prime$ are the SU(2)$_L$ singlets.

The SU(5) gauge theory contains the 24 gauge bosons and each of them
corresponds to a component of a vector superfield, ${\cal V}^A$, with
$A=1,\dots, 24$ indicating the gauge index. By exploiting the
fundamental representation of the SU(5) generators, $T^A$, we define a
$5\times 5$ matrix of the vector superfields: ${\cal V}\equiv {\cal
V}^AT^A$. The components of the matrix are written as
\begin{equation}
 {\cal V}=\frac{1}{\sqrt{2}}
\begin{pmatrix}
 \begin{matrix}
G  -\frac{2}{\sqrt{30}} B
 \end{matrix}
&
\begin{matrix}
X^{\dagger 1} \\
X^{\dagger 2} \\
X^{\dagger 3}
\end{matrix}
&
\begin{matrix}
 Y^{\dagger 1}\\ Y^{\dagger 2} \\ Y^{\dagger 3}
\end{matrix}
\\
\begin{matrix}
 X_1 & X_2 & X_3 \\
 Y_1 & Y_2 & Y_3  
\end{matrix}
&
\begin{matrix}
 \frac{1}{\sqrt{2}}W^3+\frac{3}{\sqrt{30}}B \\ W^-
\end{matrix}
&
\begin{matrix}
W^+ \\ - \frac{1}{\sqrt{2}}W^3+\frac{3}{\sqrt{30}}B
\end{matrix}
\end{pmatrix}
~,
\end{equation} 
where each component is expressed by the same symbol as that used for
the corresponding gauge field. We collectively refer to $X_\alpha$,
$Y_\alpha$, and their Hermitian conjugates as $X$ and $Y$ bosons,
and use the following notation for them:
\begin{equation}
 (X)^r_\alpha =
\begin{pmatrix}
 X^1_\alpha \\ X^2_\alpha
\end{pmatrix}
\equiv
\begin{pmatrix}
 X_\alpha \\ Y_\alpha
\end{pmatrix}~.
\end{equation}
Here $r,s,\dots$ denote the isospin indices. 

The Higgs superfields in the MSSM, on the other hand, are incorporated
into a pair of fundamental and anti-fundamental fields as
\begin{equation}
 H=
\begin{pmatrix}
 H^1_C \\ H^2_C \\ H^3_C \\ H^+_u \\ H^0_u
\end{pmatrix}
,~~~~~~\bar{H}=
\begin{pmatrix}
 \bar{H}_{C1}\\
 \bar{H}_{C2}\\
 \bar{H}_{C3}\\
 H^-_d \\ -H^0_d
\end{pmatrix}
~,
\end{equation}
where the last two components are corresponding to the MSSM Higgs superfields,
\begin{equation}
 H_u=
\begin{pmatrix}
 H^+_u \\ H^0_u
\end{pmatrix}
,~~~~~~H_d=
\begin{pmatrix}
 H^0_d \\ H^-_d
\end{pmatrix}
~,
\end{equation}
and the new Higgs superfields $H^\alpha_C$ and $\bar{H}_{C\alpha}$ are
called the Higgs color triplet superfields.
The superpotential for the Yukawa couplings of quarks and leptons is given as
\begin{align}
 W &=
\frac{1}{4}h^{ij}\epsilon_{abcde}\Psi_i^{ab} \Psi_j^{cd}H^e +\sqrt{2}
f^{ij}\Psi_i^{ab} \Phi_{ja}\bar{H}_b~,
\label{superpotential}
\end{align}
where $i,j=1,2,3$ indicate the generations and $a,b,c,\dots$
represent the SU(5) indices. 
The Yukawa couplings $h^{ij}$ and $f^{ij}$ in Eq.~(\ref{superpotential})
have redundant components and much of the degree of freedom is
eliminated through the field re-definition of $\Psi$ and $\Phi$
\cite{Ellis:1978xg}. We parametrize the couplings according to
Ref.~\cite{Hisano:1992jj} as
\begin{align}
 h^{ij}&= f_{u_i}e^{i\varphi_i}\delta_{ij}~, \\
 f^{ij}&=V^*_{ij}f_{d_j}~,
\end{align}
with $V_{ij}$ the Kobayashi-Maskawa matrix. The phase factors $\varphi_i$ are
subject to a condition:
\begin{equation}
 \varphi_1+\varphi_2+\varphi_3=0~.
\end{equation}
With the parameters, we express the matter fields in terms of the mass
eigenstates as follows:
\begin{align}
 Q^\prime_i&=
\begin{pmatrix}
 U_i \\ V_{ij}D_j
\end{pmatrix},~~~~~~L_i^\prime =
\begin{pmatrix}
 N_i \\ E_i
\end{pmatrix},
\end{align}
\begin{equation}
 \bar{U}_i^\prime =e^{-i\varphi_i}\bar{U}_i,~~~~~~
  \bar{D}_i^\prime =  \bar{D}_i,~~~~~~
\bar{E}_i^\prime=V_{ij}\bar{E}_j~.
\end{equation}
In the following discussion, we express interactions in the basis of
mass eigenstates unless otherwise noted.

Here, we do not take specific assumptions for symmetry breaking of
SU(5) and the mass generation of the colored Higgs. They are related
to suppression of the proton decay induced by dimension-five operators.

\section{Proton decay via the $X$ and $Y$ boson exchange}

Next we discuss the proton decay rate induced by the $X$ and $Y$
boson exchange. 
The couplings between the gauge bosons and the SM fermions are
given as 
\begin{align}
 {\cal L}_{\rm int}=\frac{1}{\sqrt{2}}g_5\bigl[
-\epsilon_{rs}(\overline{L^{\cal C}_i})^r\Slash{X}^s_\alpha
 P_Rd^\alpha_i
&+\epsilon_{rs}V^*_{ij}\overline{e^{\cal C}_j}\Slash{X}^r_\alpha
 P_LQ^{\prime\alpha s}\nonumber \\
&+e^{-i\varphi_i}\epsilon^{\alpha\beta\gamma}
 (\overline{Q_i})_{\alpha r} \Slash{X}^r_\beta P_L(u^{\cal C}_i)_\gamma 
+h.c.\bigr]~,
\end{align}
where $g_5$ is the SU(5) gauge coupling constant and ${\cal C}$
indicates the charge conjugation. $\epsilon_{rs}$ and
$\epsilon^{\alpha\beta\gamma}$ denote the second and third rank totally
antisymmetric tensors, respectively. 
This interaction Lagrangian causes proton decay. In the present case,
the dominant decay mode is $p\to \pi^0e^+$, and this decay process is
induced by the following effective Lagrangian:
\begin{align}
 {\cal L}_{\rm
 eff}&=-\frac{g_5^2}{M_X^2}e^{i\varphi_1}\epsilon_{\alpha\beta\gamma}
 \bigl[A_R^{(1)}(\overline{u^c})^\alpha
 P_Rd^\beta\overline{e^+}P_Lu^\gamma 
+A^{(2)}_R
(1+|V_{ud}|^2)(\overline{u^c})^\alpha P_L d^\beta
 \overline{e^+}P_Ru^\gamma \bigr]~.
\label{effective_lagrangian}
\end{align}
where $M_X$ is the mass of $X$ and $Y$ bosons and $P_{L/R}\equiv
\frac{1}{2}(1\mp \gamma_5)$.
The renormalization effects resulting from the anomalous dimensions of
the operators are represented by $A_R^{(1)}$ and $A_R^{(2)}$, which we
will evaluate below. 
The hadron matrix elements of the operators are evaluated by using the
ordinary chiral Lagrangian method\footnote{
Calculation of the matrix elements is also conducted by using the direct
method \cite{Aoki:2006ib}. Recent progress, in which the quenched
approximation is not used, is reported in Ref.~\cite{GUT2012}.
}~\cite{Claudson:1981gh}:
\begin{equation}
 \langle \pi^0 | \epsilon_{\alpha\beta\gamma}
[(\overline{u^c})^\alpha P_Rd^\beta]P_Lu^\gamma 
|p(\bm{p},s)\rangle
=
\frac{\alpha_{\rm H}}{\sqrt{2}f_\pi}(1+D+F)P_Lu_p(\bm{p},s)~,
\end{equation}
and
\begin{equation}
  \langle \pi^0 | \epsilon_{\alpha\beta\gamma}
[(\overline{u^c})^\alpha P_Ld^\beta]P_Ru^\gamma 
|p(\bm{p},s)\rangle
=
\frac{\alpha_{\rm H}}{\sqrt{2}f_\pi}(1+D+F)P_Ru_p(\bm{p},s)~,
\end{equation}
where $f_\pi$ is the pion decay constant, and
$\alpha_{\rm H}$ is defined by the equation
\begin{equation}
  \langle 0 | \epsilon_{\alpha\beta\gamma}
[u^\alpha C P_Rd^\beta]P_Lu^\gamma 
|p(\bm{p},s)\rangle
=
\alpha_{\rm H}P_Lu_p(\bm{p},s)~,
\end{equation}
with $C$ the charge conjugation matrix. The value of $\alpha_{\rm H}$ is
computed in Ref.~\cite{Aoki:2008ku} as
\begin{equation}
 \alpha_{\rm H} (2 ~{\rm GeV})
=-0.0112\pm 0.0012_{(\rm stat)} \pm 0.0022_{(\rm syst)}
~{\rm GeV}^3,
\end{equation}
at the renormalization scale $\mu=2$ GeV. By using the matrix elements,
we obtain the partial decay width $\Gamma (p\rightarrow \pi^0 e^+)$
induced by the effective Lagrangian in Eq.~(\ref{effective_lagrangian}):
\begin{align}
\Gamma (p\rightarrow \pi^0 e^+)
= \frac{\pi}{4}\frac{\alpha_5^2}{M_X^4}\frac{m_p}{f_\pi^2}
 \alpha_{\rm H}^2|1+D+F|^2\biggl(1-\frac{m_\pi^2}{m_p^2}\biggr)^2
 \bigl[
\bigl(A^{(1)}_R\bigr)^2+\bigl(A^{(2)}_R\bigr)^2(1+|V_{ud}|^2)^2
\bigr]~,
\label{decay_width}
\end{align}
where $m_p$ and $m_\pi$ are the masses of proton and the neutral pion,
respectively, and $\alpha_5\equiv g_5^2/4\pi$~. In the following
calculation, we take $f_\pi=130~{\rm MeV}$, $D=0.80$, and $F=0.47$ as in
Ref.~\cite{Aoki:2008ku}.

Now, in order to evaluate the proton decay rate, all we have to do is to
determine the unified gauge coupling constant $\alpha_5$, the $X$ boson
mass $M_X$, and the renormalization factors $A^{(1)}_R$ and
$A^{(2)}_R$. They are dependent on each GUT model.

The unified gauge coupling constant $\alpha_5$ is computed by solving
the renormalization group equations (RGEs) for the gauge
coupling constants $g_a$ ($a=1,2,3$) in the SM gauge interactions. 
In this article, we consider the gauge
coupling running up to the two-loop level, and exploit the
$\overline{\rm DR}$ renormalization scheme in order to respect the
supersymmetry \cite{Siegel:1979wq}.
Furthermore, we adopt the definition of $\alpha_5$ as $\alpha_5\equiv
g_3^2(M_{\rm GUT})/4\pi$ with $M_{\rm GUT}=1.5\times 10^{16}$~GeV.

In the MSSM, the two-loop
renormalization group equations (RGEs) for the gauge coupling constants are
given as \cite{Bjorkman:1985mi}
\begin{equation}
 \mu \frac{\partial g_a}{\partial \mu}=\frac{1}{16\pi^2}b_a 
^{(1)}g^3_a
+\frac{g_a^3}{(16\pi^2)^2}\biggl[
\sum_{b=1}^{3}b_{ab}^{(2)}g_b^2 -\sum_{i=t,b,\tau}c_{ai}~ y^2_i 
\biggr]~,
\end{equation}
where
\begin{equation}
b^{(1)}_a=
\begin{pmatrix}
 33/5 \\ 1 \\ -3
\end{pmatrix}
~,~~~~b_{ab}^{(2)}=
\begin{pmatrix}
 199/25 & 27/5 & 88/5 \\
 9/5 & 25 & 24 \\
 11/5 & 9 & 14 
\end{pmatrix} 
~,~~~~~(a,b=1,2,3)
\label{MSSM_beta}
\end{equation}
and
\begin{equation}
 c_{ai}=
\begin{pmatrix}
 26/5 & 14/5 & 18/5 \\
 6 & 6 & 2 \\
 4 & 4 & 0
\end{pmatrix}
~, ~~~~~~(i=t,b,\tau)
\end{equation}
with $y_i$ the Yukawa couplings. Here, we use the SU(5) normalization
for the U(1) hypercharge. Since the Yukawa
couplings enter into the two-loop level contributions to the gauge
coupling RGEs, it is sufficient to consider the RGEs for the Yukawa
couplings at one-loop level. They are given as 
\begin{align}
 \mu\frac{\partial}{\partial \mu}y_t&=\frac{1}{16\pi^2}y_t
\biggl[6y_t^2+y^2_b -\frac{13}{15}g^2_1-3g_2^2-\frac{16}{3}
 g_3^2\biggr] , \nonumber \\
 \mu\frac{\partial}{\partial \mu}y_b &=\frac{1}{16\pi^2}y_b
\biggl[6y_b^2+y_t^2+y_\tau^2
 -\frac{7}{15}g^2_1-3g_2^2-\frac{16}{3}
 g_3^2\biggr] , \nonumber \\
 \mu\frac{\partial}{\partial \mu}y_\tau &=\frac{1}{16\pi^2}y_\tau
\biggl[3y_b^2+4y_\tau^2 
 -\frac{9}{5}g^2_1-3g_2^2\biggr] .
\end{align}

In the SM, on the other hand, the coefficients for the gauge coupling
beta functions are~\cite{Machacek:1983tz}
\begin{equation}
 b_a^{(1)}=
\begin{pmatrix}
 41/10 \\ -19/6 \\ -7
\end{pmatrix}
,~~~~~~b^{(2)}_{ab}=
\begin{pmatrix}
 199/50 & 27/10 & 44/5 \\
 9/10 & 35/6 & 12 \\
 11/10 & 9/2 & -26
\end{pmatrix}
~,
\end{equation}
and
\begin{equation}
 c_{ai}=
\begin{pmatrix}
 17/10 & 1/2 & 3/2 \\
 3/2 & 3/2 & 1/2 \\
 2 & 2 & 0
\end{pmatrix}
~.
\end{equation}
The running of the Yukawa couplings in this case is given as follows:
\begin{align}
 \mu\frac{\partial}{\partial \mu}y_t&=\frac{1}{16\pi^2}y_t
\biggl[\frac{9}{2}y_t^2 +\frac{3}{2}y_b^2 +y_\tau^2
 -\frac{17}{20}g^2_1-\frac{9}{4}g_2^2-8
 g_3^2\biggr] , \nonumber \\
 \mu\frac{\partial}{\partial \mu}y_b&=\frac{1}{16\pi^2}y_b
\biggl[\frac{3}{2}y_t^2+\frac{9}{2}y_b^2 +y_\tau^2
 -\frac{1}{4}g^2_1-\frac{9}{4}g_2^2-8
 g_3^2\biggr] , \nonumber \\ 
 \mu\frac{\partial}{\partial \mu}y_\tau&=\frac{1}{16\pi^2}y_\tau
\biggl[3y_t^2 +3y_b^2+\frac{5}{2}y_\tau^2
 -\frac{9}{4}g^2_1-\frac{9}{4}g_2^2\biggr] .
\end{align}
Modifications in the coefficients in particular models are mentioned to
in the following sections.

Now we deal with the renormalization factors, $A^{(1)}_R$ and
$A^{(2)}_R$. The factors are expressed as the product of the long- and
short-distance factors, {\it i.e.}, 
\begin{equation}
 A^{(i)}_R=A_L\cdot A_S^{(i)},~~~~~~(i=1,2)
\label{AR}
\end{equation}
where $A_L$ and $A_S^{(i)}$ represent the long- and short-distance
factors, respectively. The long-distance contribution $A_L$ is common to
$A_R^{(1)}$ and $A_R^{(2)}$, and independent of the high-energy
physics. Its value is evaluated as
\begin{equation}
 A_L=1.25~,
\end{equation}
at two-loop level. Details of the calculation are given in Appendix.

The short-distance factors are, on the other hand, model-dependent
quantities. For instance, if there is no threshold between the
electroweak and the GUT scales they are evaluated at one-loop level in
Refs.~\cite{Abbott:1980zj, Munoz:1986kq} as follows:
\begin{align}
 A_S^{(1)}&=\biggl[
\frac{\alpha_3(m_Z)}{\alpha_3(M_{\rm GUT})}
\biggr]^{-\frac{\gamma_3}{b_3^{(1)}}}\biggl[
\frac{\alpha_2(m_Z)}{\alpha_2(M_{\rm GUT})}
\biggr]^{-\frac{\gamma_2}{b_2^{(1)}}}\biggl[
\frac{\alpha_1(m_Z)}{\alpha_1(M_{\rm GUT})}
\biggr]^{-\frac{\gamma_1^{(1)}}{b_1^{(1)}}}~, \nonumber \\
 A_S^{(2)}&=\biggl[
\frac{\alpha_3(m_Z)}{\alpha_3(M_{\rm GUT})}
\biggr]^{-\frac{\gamma_3}{b_3^{(1)}}}\biggl[
\frac{\alpha_2(m_Z)}{\alpha_2(M_{\rm GUT})}
\biggr]^{-\frac{\gamma_2}{b_2^{(1)}}}\biggl[
\frac{\alpha_1(m_Z)}{\alpha_1(M_{\rm GUT})}
\biggr]^{-\frac{\gamma_1^{(2)}}{b_1^{(1)}}}~,
\end{align}
where, in the MSSM,
\begin{equation}
 \gamma_3=\frac{4}{3},~~~~\gamma_2=\frac{3}{2},~~~~
  \gamma_1^{(1)}=\frac{11}{30},~~~~\gamma_1^{(2)}=\frac{23}{30}~,
\end{equation}
while in the SM,
\begin{equation}
 \gamma_3=2,~~~~\gamma_2=\frac{9}{4},~~~~
  \gamma_1^{(1)}=\frac{11}{20},~~~~\gamma_1^{(2)}=\frac{23}{20}~.
\end{equation}
The extension of the result to each case is referred to in the subsequent
sections.

For reference, we evaluate the proton-decay lifetime assuming the SUSY
scale to be 1~TeV, {\it i.e.}, all the superparticles are assumed to
have masses of ${\cal O}(1)$~TeV. The result is
\begin{equation}
 \tau(p\to e^+ \pi^0)= 1.16\times
  10^{35}\times\biggl(\frac{M_X}{1.0\times 10^{16}~{\rm GeV}}\biggr)^4
~~~~~{\rm years}~.
\label{lifetime5_0_10_0}
\end{equation}
Here we neglect the possible effects of the threshold corrections
from particles whose masses are around the GUT scale. We also neglect
them in the following calculations, since the effects are completely
model-dependent.

\section{Proton decay with vector-like matters}
\label{vectorlike}

In this section, we discuss the grand unified models in which extra
vector-like matters are added into the SUSY SU(5) GUTs. We assume that
there exist $n_5$ and $n_{10}$ pairs of chiral supermultiplets which
transform as ${\bf 5}+\overline{\bf 5}$ and ${\bf 10}+\overline{\bf
  10}$ representations, respectively, and evaluate the proton decay
rate for the cases.  For brevity, all of the multiplets are assumed to
have the same mass, $M$. As mentioned to in the Introduction, such
models are motivated by the symptoms of Higgs boson with its
relatively heavy mass, $m_h \approx 125$~GeV, reported recently at the
LHC \cite{:2012si}; if the additional multiplets with mass around
weak scale couple to the MSSM Higgs bosons, the Higgs mass $m_h$ can
be raised on account of the quantum corrections \cite{Moroi:1992zk}.
The gauge mediation requires the vector-like matters as the 
SUSY-breaking messenger. Thus, we take a wide parameter range for $M$.

The existence of the extra vector-like multiplets modifies the
RGEs for the gauge coupling constants, as well as the short-distance
renormalization factors. 
The beta functions of the gauge coupling constants in the MSSM receive
additional contributions \cite{Ghilencea:1997mu}:
\begin{align}
 \delta b^{(1)}_a&=
\begin{pmatrix}
 n_5+3n_{10} \\
 n_5+3n_{10} \\
 n_5+3n_{10} 
\end{pmatrix}
, \nonumber\\
 \nonumber\\
\delta b^{(2)}_{ab} &=
\begin{pmatrix}
 \frac{7}{15}n_5+\frac{23}{5}n_{10}&\frac{9}{5}n_5+\frac{3}{5}n_{10}
&\frac{32}{15}n_5+\frac{48}{5}n_{10} \\ 
\\
\frac{3}{5}n_5+\frac{1}{5}n_{10} &7n_5+21n_{10}&16n_{10}\\
\\
\frac{4}{15}n_5+\frac{6}{5}n_{10}&6n_{10}& \frac{34}{3}n_5+34n_{10}
\end{pmatrix}
.
\label{beta_vec}
\end{align}
Here $\delta b^{(1)}_a$ and $\delta b^{(2)}_{ab}$ are the corrections
to the leading and the next-to-leading order contributions for the
beta functions of the gauge coupling constants, $b^{(1)}_a$ and
$b^{(2)}_{ab}$, which are defined in Eq.~(\ref{MSSM_beta}).  Here, we
ignore the contribution from the Yukawa couplings of the vector-like
matters at two-loop level for simplicity.  From the one-loop
contribution in Eq.~(\ref{beta_vec}), it is found that addition of
vector-like matters equally changes the running of the gauge
couplings, and thus maintains the gauge coupling unification.

The corrections to the one-loop contribution, $\delta b^{(1)}_a$, also
modify $A_S^{(1)}$ and $A_S^{(2)}$. To be concrete, the renormalization
effects of the energy scale above $M$ are given as
\begin{align}
 A_S^{(1)}&=\biggl[
\frac{\alpha_3(M)}{\alpha_3(M_{\rm GUT})}
\biggr]^{-\frac{4}{3 [b_3^{(1)}+ \delta b_3^{(1)}]}}\biggl[
\frac{\alpha_2(M)}{\alpha_2(M_{\rm GUT})}
\biggr]^{-\frac{3}{2[b_2^{(1)}+\delta b_2^{(1)}]}}\biggl[
\frac{\alpha_1(M)}{\alpha_1(M_{\rm GUT})}
\biggr]^{-\frac{11}{30[b_1^{(1)}+\delta b_1^{(1)}]}}~, \nonumber \\
 A_S^{(2)}&=\biggl[
\frac{\alpha_3(M)}{\alpha_3(M_{\rm GUT})}
\biggr]^{-\frac{4}{3 [b_3^{(1)}+ \delta b_3^{(1)}]}}\biggl[
\frac{\alpha_2(M)}{\alpha_2(M_{\rm GUT})}
\biggr]^{-\frac{3}{2[b_2^{(1)}+\delta b_2^{(1)}]}}\biggl[
\frac{\alpha_1(M)}{\alpha_1(M_{\rm GUT})}
\biggr]^{-\frac{23}{30[b_1^{(1)}+\delta b_1^{(1)}]}}~.
\end{align}

These modifications result in alternations of the proton-decay lifetime.
In order to parametrize the alternations, we define the following ratio:
\begin{equation}
 R\equiv \frac{\tau(p\to e^+\pi^0)|_{\rm w/}}{\tau(p\to e^+\pi^0)|_{\rm
  w/o}}~, 
\end{equation}
where $\tau(p\to e^+\pi^0)|_{\rm w/(w/o)}$ represents the proton-decay
lifetime with (without) vector-like matters. As seen form
Eq.~(\ref{decay_width}), this ratio does not depend on the $X$-boson
mass but only on $\alpha_5$, $A_S^{(1)}$ and $A_S^{(2)}$:
\begin{equation}
 R=\frac{\alpha_5^2\bigl[\bigl(A^{(1)}_R\bigr)^2
+\bigl(A^{(2)}_R\bigr)^2(1+|V_{ud}|^2)^2 \bigr]_{\rm w/o}}
{\alpha_5^2\bigl[\bigl(A^{(1)}_R\bigr)^2
+\bigl(A^{(2)}_R\bigr)^2(1+|V_{ud}|^2)^2 \bigr]_{\rm w/}}~.
\end{equation}
Here, w/ (w/o) again implies that the factor is for the case with
(without) vector-like matters.

\begin{figure}[t]
\begin{center}
\includegraphics[height=65mm]{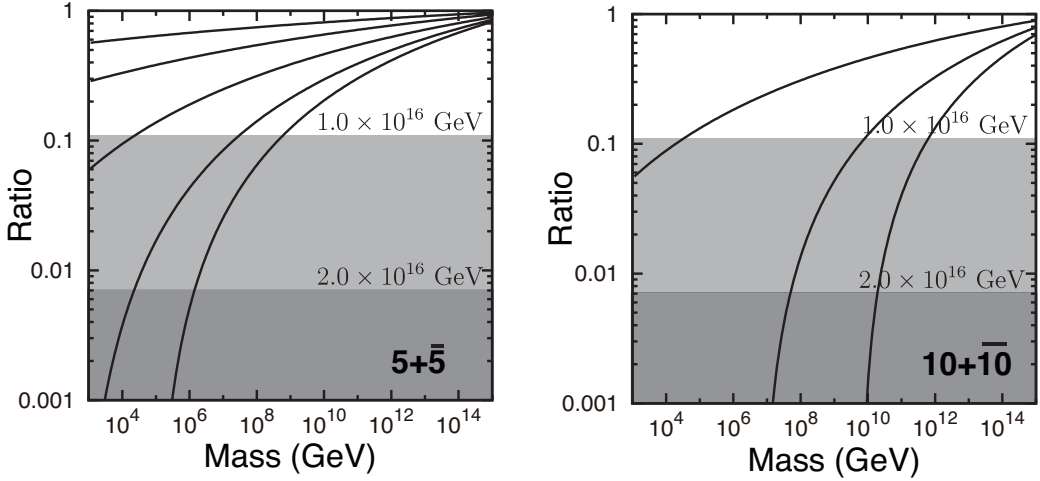}
\caption{Ratio $R$ as functions of vector-like matter
 mass. Left and right graphs show $n_5=1,\dots, 5$ and $n_{10}=1,\dots
 3$ in solid lines from top to bottom, respectively. 
 Light (dark) shaded region is excluded by the
 current experimental limit, $\tau(p\to e^+ \pi^0)> 1.29\times
 10^{34}$~years at 90\% confidence level \cite{Nishino:2012rv}, in the case
 of $M_X=1.0\times 10^{16}$~GeV ($2.0\times 10^{16}$~GeV). SUSY
 scale is set to be 1~TeV.}
\label{fig:vectorlike}
\end{center}
\end{figure}

\begin{figure}[t]
\begin{center}
\includegraphics[height=65mm]{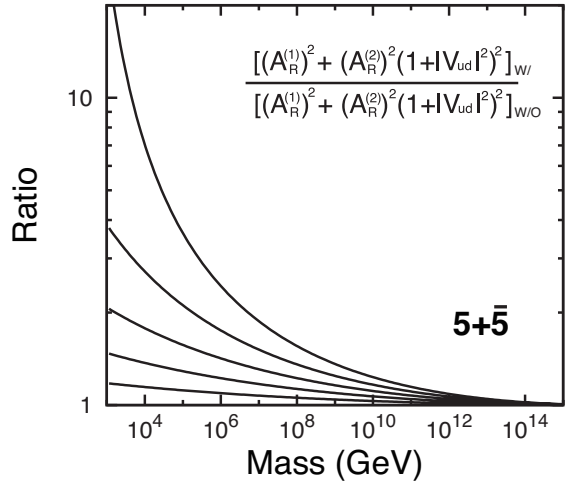}
\caption{Enhancement of renormalization factors as functions of
 vector-like matter mass. We take $n_5=1,\dots, 5$ from
 bottom to top. SUSY scale is set to be 1~TeV.}
\label{fig:renormalization}
\end{center}
\end{figure}

In Fig.~\ref{fig:vectorlike}, we plot the ratio $R$ as
functions of the masses of the vector-like multiplets. In the
analysis, the SUSY scale is set to be $1$~TeV.  Each solid line in the left
graph corresponds to the number of ${\bf 5}+\overline{\bf 5}$
multiplets $n_5=1, 2, \dots, 5$ from top to bottom without ${\bf
  10}+\overline{\bf 10}$ multiplets. In the right graph, on the other
hand, the solid lines represent the cases with ${\bf 10}+\overline{\bf
  10}$ multiplets ($n_{10}=1,2,3$, from top to bottom) and
$n_5=0$. 
The light (dark) shaded region is excluded by the current
experimental limit, $\tau(p\to e^+ \pi^0)> 1.29\times 10^{34}$~years at
90\% confidence level \cite{Nishino:2012rv}  in the case of $M_X=1.0\times
10^{16}$~GeV ($2.0\times 10^{16}$~GeV). On the whole parameter
region in this figure, the unified 
gauge coupling constant, $\alpha_5$, is less than $0.6$, thus,
perturbativity is maintained.
In order to clarify the effects of the renormalization factors on the
enhancement of proton decay rate, we also plot the ratio of the factors,
$[(A^{(1)}_R)^2 +(A^{(2)}_R)^2(1+|V_{ud}|^2)^2]_{\rm w/}/ [(A^{(1)}_R)^2
+(A^{(2)}_R)^2(1+|V_{ud}|^2)^2 ]_{\rm w/o}$, as functions of the masses
of the vector-like multiplets in Fig.~\ref{fig:renormalization}. Here we 
set $n_5=1,\dots, 5$ from bottom to top and the SUSY scale to be
1~TeV. The renormalization factor in the absence of the vector-like
multiplets is calculated as $[(A^{(1)}_R)^2 +(A^{(2)}_R)^2(1+|V_{ud}|^2)^2
]_{\rm w/o}=40$ .
These results indicate
that the MSSM with vector-like matters might cause proton decay fast
enough to be detected by the current or future experiments. In
particular, if there are three ${\bf 5}+\overline{\bf 5}$ vector-like
matters or is a ${\bf 10}+\overline{\bf 10}$ multiplet at the TeV
scale, the SUSY SU(5) GUTs are excluded even by the present limit for
$M_X=1.0\times 10^{16}$ GeV.

\begin{figure}[t]
\begin{center}
\includegraphics[height=70mm]{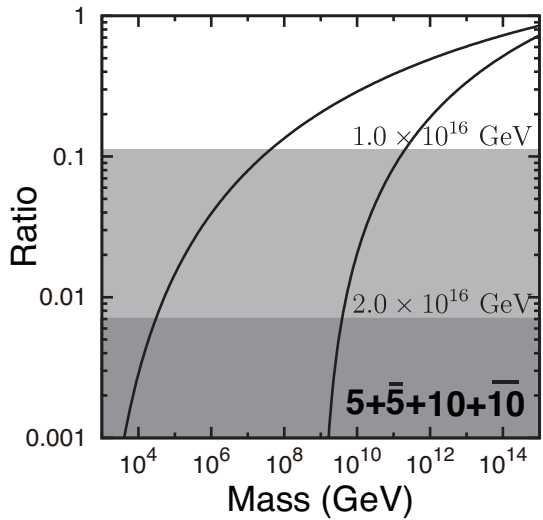}
\caption{Ratio $R$ as functions of the masses of the
 multiplets. Solid lines represent $n_g=1,2$ multiplets of ${\bf
 5}+\overline{\bf 5}+{\bf 10}+\overline{\bf 10}$ representations from
 top to bottom. Light (dark) shaded region is excluded by the
 current experimental limit, $\tau(p\to e^+ \pi^0)> 1.29\times
 10^{34}$~years at 90\% confidence level \cite{Nishino:2012rv}  in the case
 of $M_X=1.0\times 10^{16}$~GeV ($2.0\times 10^{16}$~GeV). SUSY scale is
 set to be 1~TeV.} 
\label{fig:gen}
\end{center}
\end{figure}
In addition, we calculate the ratio in the case where
$n_g$ pairs of ${\bf 5}+\overline{\bf 5}+{\bf 10}+\overline{\bf 10}$
multiplets are introduced. This set of representations is the same as
those of one generation of chiral matter fields and their corresponding
complex representations. In this case, the beta functions are
equal to those of $n_5=n_{10}=n_g$ in Eq.~(\ref{beta_vec}). After
carrying out a similar calculation, we plot the
ratio $R$ against the masses of the multiplets in
Fig.~\ref{fig:gen}. Here, solid lines represent $n_g=1,2$ multiplets
of ${\bf 5}+ \overline{\bf 5}+{\bf 10}+\overline{\bf 10}$ from top to
bottom. We again set the SUSY scale to be 1~TeV. 
From this figure, it is found that a pair
of extra generations which have masses below $10^{7}$~GeV yield too
large proton decay rate to be excluded by the current experimental limit
in the case of $M_X=1.0\times 10^{16}$~GeV.
Although the threshold corrections around the GUT scale might
alter the predicted values of proton lifetime, a growing tendency in the
proton decay rate is independent of particular models. Therefore, the
proton-decay experiments are extremely promising for constraining such
scenario.

\section{Proton decay with heavy SUSY particles}
\label{heavy_susy}

In turn, we discuss the SUSY GUT with sfermions
having masses much larger than the TeV scale. Such a mass
spectrum might be realized when the supersymmetry is broken via the
anomaly-mediation mechanism \cite{Randall:1998uk}, and there have been a
lot of works which examine the scenario \cite{ArkaniHamed:2004fb}. In
this scenario, although a high degree of fine-tuning is inevitable, the
SUSY flavor and CP problems are relaxed owing to heavy masses of
sfermions \cite{Gabbiani:1996hi}. The thermal relic scenario of dark
matter in the Universe is still achieved~\cite{Hisano:2006nn}, and the
dark matter is to be directly detected in the future experiments
\cite{Hisano:2010fy}. 
In addition, this heavy scale SUSY scenario is also suggested by the
recent LHC results since there has been no signal of superparticles and
the ATLAS and CMS collaborations provide stringent limits on the masses
of colored particles \cite{Aad:2011ib}. Moreover, in this case
the SM-like Higgs boson mass easily reaches $\sim 125$ GeV as the
radiative corrections are enhanced due to the heavy colored particles.

In the following discussion, we assume that the masses of squarks and
sleptons are around the scale of $M_{\rm SF}$, which is much
higher then the electroweak scale, while the masses of gauginos,
higgsinos, and Higgs bosons except the lightest one are ${\cal
O}(1)$~TeV, which is denoted by $M_{\rm GH}$ hereafter. 
Therefore, between $M_{\rm GH}$ and $M_{\rm SF}$, the latter fields are
to be added to the RGE analysis. Furthermore, we assume that their
contributions to the gauge-coupling running through the Yukawa
couplings, as well as those with sfermions running in the loops, are
negligible. Hence, the beta functions of the gauge couplings in this region 
are given as
\begin{equation}
 b_a^{(1)}=
\begin{pmatrix}
 23/5 \\ -1 \\ -5
\end{pmatrix}
,~~~~~~b^{(2)}_{ab}=
\begin{pmatrix}
 104/25 & 18/5 & 44/5 \\
 6/5 & 32 & 12 \\
 11/10 & 9/2 & 22
\end{pmatrix}
~,
\end{equation}
with $c_{ai}$ the same as the SM ones.
Since each one-loop contribution $b^{(1)}_a$ in the above equation
differ from that in Eq.~(\ref{MSSM_beta}) by the same number (2 in this
case), the perturbative gauge coupling unification is again preserved in
the present case.

\begin{figure}[t]
\begin{center}
\includegraphics[height=70mm]{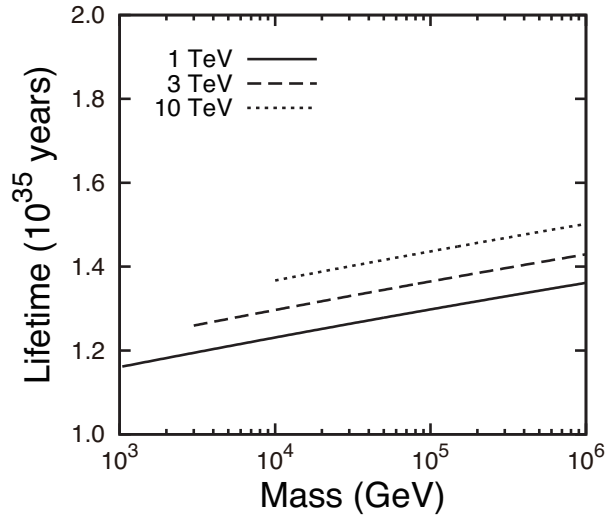}
\caption{Proton lifetime against $M_{\rm SF}$. Solid, dashed, and
 dotted lines correspond to the cases where $M_{\rm GH}$ is set to be 1,
 3, and 10~TeV, respectively. $X$-boson mass is taken to be $1.0\times
 10^{16}$~GeV. }  
\label{fig:heavysusy}
\end{center}
\end{figure}

In Fig.~\ref{fig:heavysusy}, we plot the lifetime of proton against
$M_{\rm SF}$. Solid, dashed, and dotted lines correspond to the
cases where $M_{\rm GH}$ is set to be 1, 3, and 10~TeV, respectively. 
The $X$-boson mass is taken to be $1.0\times 10^{16}$~GeV in this
figure. It is found that the proton-decay lifetime is slightly extended
, although the enhancement factor is less than two. 
The contribution of the renormalization factors to the alternation
of the proton lifetime is less significant in this case than that in the
case discussed in Sec.~\ref{vectorlike}; It is at most a few \% on the
whole parameter region in this figure.
Compared with the
case discussed in the previous section, the present situation does not
so much change the proton decay rate. Hence, searching for the proton
decay is still stimulating even for the heavy SUSY scenario.

\section{Conclusion}
\label{conclusion}

We have studied the proton decay rate under the two situations; one
is the MSSM with vector-like multiplets, and the other is the MSSM with
heavy sfermions. It is found that the proton lifetime is
significantly reduced in the former case, while in the latter case it is
slightly prolonged. In any case, the proton-decay experiments are,
together with the LHC experiment and other precision measurements,
expected to shed light on the supersymmetric grand unified models, as
well as the supersymmetry itself.

\section*{Acknowledgments}

We thank Yasumichi Aoki for useful discussion. This work is supported
by Grant-in-Aid for Scientific research from the Ministry of
Education, Science, Sports, and Culture (MEXT), Japan, No. 20244037,
No. 20540252, No. 22244021 and No.23104011 (JH), and also by World
Premier International Research Center Initiative (WPI Initiative),
MEXT, Japan.
The work of NN is supported by
Research Fellowships of the Japan Society for the Promotion of Science
for Young Scientists.

\section*{Appendix: Long-distance part of the renormalization factors}
\appendix

In this Appendix, we demonstrate the evaluation of the long-distance
contribution to the renormalization factors, {\it e.g.}, $A_L$ in
Eq.~(\ref{AR}). First, we write down the two-loop renormalization group
equations for the strong coupling constant below the electroweak scale:
\begin{equation}
 \mu \frac{\partial g_s}{\partial
  \mu}=\frac{1}{16\pi^2}b_1g_s^3+\frac{1}{(16\pi^2)^2} b_2g_s^5~,
\end{equation} 
with $g_s$ the strong coupling constant and 
\begin{equation}
 b_1=-\biggl(11-\frac{2}{3}N_f\biggr),~~~~~
  b_2=-\biggl(102-\frac{38}{3}N_f\biggr)~,
\label{b1b2}
\end{equation}
where $N_f$ denotes the number of quark flavors in an effective theory.

The long-distance factor, $A_L$, is determined by the ratio of the
coefficients for the effective operators at the scale of $m_Z$ and 2~GeV:
\begin{equation}
 A_L\equiv \frac{C(2 ~{\rm GeV})}{C(m_Z)}~,
\end{equation}
with the coefficient $C(\mu)$ satisfying the following RGE at two-loop
level~\cite{Nihei:1994tx}:
\begin{equation}
 \mu \frac{\partial}{\partial \mu}C(\mu)=
-\biggl[
4\frac{\alpha_s}{4\pi}+\biggl(\frac{4}{3}+\frac{4}{9}N_f
\biggr)\frac{\alpha_s^2}{(4\pi)^2}
\biggr]C(\mu)~.
\end{equation}
The solution of the equation is
\begin{equation}
 \frac{C(\mu)}{C(\mu_0)}=\biggl[
\frac{\alpha_s(\mu)}{\alpha_s(\mu_0)}
\biggr]^{-\frac{2}{b_1}}
\biggl[
\frac{4\pi b_1+b_2\alpha_s(\mu)}
{4\pi b_1+b_2\alpha_s(\mu_0)}
\biggr]^{\bigl(\frac{2}{b_1}-\frac{6+2N_f}{9 b_2}\bigr)}~,
\end{equation}
with $b_1$ and $b_2$ given in Eq.~(\ref{b1b2}). Thus, $A_L$ is given as
\begin{align}
 A_L=\biggl[
\frac{\alpha_s(2~{\rm GeV})}{\alpha_s(m_b)}
\biggr]^{\frac{6}{25}}\biggl[
\frac{\alpha_s(m_b)}{\alpha_s(m_Z)}
\biggr]^{\frac{6}{23}}
\biggl[
\frac{\alpha_s(2~{\rm GeV})+\frac{50\pi}{77}}
{\alpha_s(m_b)+\frac{50\pi}{77}}
\biggr]^{-\frac{173}{825}}
\biggl[
\frac{\alpha_s(m_b)+\frac{23\pi}{29}}
{\alpha_s(m_Z)+\frac{23\pi}{29}}
\biggr]^{-\frac{430}{2001}}~,
\end{align}
and numerically it turns out to be
\begin{equation}
 A_L = 1.25~.
\end{equation}

{}


\begin{thebibliography}{99}

\bibitem{Georgi:1974sy} 
  H.~Georgi and S.~L.~Glashow,
  Phys.\ Rev.\ Lett.\  {\bf 32}, 438 (1974).

\bibitem{Dimopoulos:1981yj} 
  S.~Dimopoulos, S.~Raby and F.~Wilczek,
  Phys.\ Rev.\ D {\bf 24}, 1681 (1981);\\
%
W.~Marciano and G.~Senjanovi\' c,
 Phys. Rev. D {\bf 25}, 3092 (1982);\\
M.B.~Einhorn and D.R.~Jones,
Nucl. Phys. B {\bf 196}, 475 (1982);\\
  U.~Amaldi, W.~de Boer and H.~Furstenau,
  Phys.\ Lett.\ B {\bf 260}, 447 (1991);\\
%
  P.~Langacker and M.~-x.~Luo,
  Phys.\ Rev.\ D {\bf 44}, 817 (1991).

\bibitem{Georgi:1974yf} 
  H.~Georgi, H.~R.~Quinn and S.~Weinberg,
  Phys.\ Rev.\ Lett.\  {\bf 33}, 451 (1974);\\
%
  S.~Dimopoulos and S.~Raby,
  Nucl.\ Phys.\ B {\bf 192}, 353 (1981);\\
%
  E.~Witten,
  Nucl.\ Phys.\ B {\bf 188}, 513 (1981).

\bibitem{Dimopoulos:1981zb} 
  S.~Dimopoulos and H.~Georgi,
  Nucl.\ Phys.\ B {\bf 193}, 150 (1981);\\
%
  N.~Sakai,
  Z.\ Phys.\ C {\bf 11}, 153 (1981).

\bibitem{Sakai:1981pk} 
  N.~Sakai and T.~Yanagida,
  Nucl.\ Phys.\ B {\bf 197}, 533 (1982);\\
%
  S.~Weinberg,
  Phys.\ Rev.\ D {\bf 26}, 287 (1982).

\bibitem{Nishino:2012rv} 
  H.~Nishino, K.~Abe, Y.~Hayato, T.~Iida, M.~Ikeda, J.~Kameda, Y.~Koshio and M.~Miura {\it et al.},
  arXiv:1203.4030 [hep-ex].

\bibitem{Miura:2010zz} 
  M.~Miura,
  PoS ICHEP {\bf 2010}, 408 (2010).

\bibitem{Goto:1998qg} 
  T.~Goto and T.~Nihei,
  Phys.\ Rev.\ D {\bf 59}, 115009 (1999);\\
  H.~Murayama and A.~Pierce,
  Phys.\ Rev.\ D {\bf 65}, 055009 (2002)
.

\bibitem{Hisano:1992ne} 
  J.~Hisano, H.~Murayama and T.~Yanagida,
  Phys.\ Lett.\ B {\bf 291}, 263 (1992);\\
%
  K.~S.~Babu and S.~M.~Barr,
  Phys.\ Rev.\ D {\bf 48}, 5354 (1993);\\
  J.~Hisano, T.~Moroi, K.~Tobe and T.~Yanagida,
  Phys.\ Lett.\ B {\bf 342}, 138 (1995);\\
%
%
  B.~Bajc, P.~Fileviez Perez and G.~Senjanovic,
  Phys.\ Rev.\ D {\bf 66}, 075005 (2002);\\
  B.~Bajc, P.~Fileviez Perez and G.~Senjanovic,
  hep-ph/0210374.
\bibitem{Peccei:1977hh} 
  R.~D.~Peccei and H.~R.~Quinn,
  Phys.\ Rev.\ Lett.\  {\bf 38}, 1440 (1977).

\bibitem{Hall:2001pg} 
  L.~J.~Hall and Y.~Nomura,
  Phys.\ Rev.\ D {\bf 64}, 055003 (2001)
.

\bibitem{Aad:2011ib}
  G.~Aad {\it et al.}  [ATLAS Collaboration],
  arXiv:1109.6572 [hep-ex];\\
%
 S.~Chatrchyan {\it et al.} [ CMS Collaboration ],
 [arXiv:1109.2352 [hep-ex]].


\bibitem{:2012si} 
  [ATLAS Collaboration],
  arXiv:1202.1408 [hep-ex];\\
%
  S.~Chatrchyan {\it et al.}  [CMS Collaboration],
  arXiv:1202.1488 [hep-ex].

\bibitem{Ibe:2011aa} 
  M.~Ibe and T.~T.~Yanagida,
  Phys.\ Lett.\ B {\bf 709}, 374 (2012);\\
  M.~Ibe, S.~Matsumoto and T.~T.~Yanagida,
  Phys.\ Rev.\ D {\bf 85}, 095011 (2012)
.

\bibitem{Ajaib:2012vc} 
  M.~A.~Ajaib, I.~Gogoladze, F.~Nasir and Q.~Shafi,
  arXiv:1204.2856 [hep-ph].

\bibitem{Moroi:1992zk} 
  T.~Moroi and Y.~Okada,
  Phys.\ Lett.\ B {\bf 295}, 73 (1992);\\
%
  K.~S.~Babu, I.~Gogoladze, M.~U.~Rehman and Q.~Shafi,
  Phys.\ Rev.\ D {\bf 78}, 055017 (2008);\\
  S.~P.~Martin,
  Phys.\ Rev.\ D {\bf 81}, 035004 (2010);\\
  M.~Asano, T.~Moroi, R.~Sato and T.~T.~Yanagida,
  Phys.\ Lett.\ B {\bf 705}, 337 (2011);\\
  M.~Endo, K.~Hamaguchi, S.~Iwamoto and N.~Yokozaki,
  Phys.\ Rev.\ D {\bf 84}, 075017 (2011); \\
  J.~L.~Evans, M.~Ibe and T.~T.~Yanagida,
  arXiv:1108.3437 [hep-ph].

\bibitem{Hebecker:2002rc} 
  A.~Hebecker and J.~March-Russell,
  Phys.\ Lett.\ B {\bf 539}, 119 (2002);
  L.~J.~Hall and Y.~Nomura,
  Phys.\ Rev.\ D {\bf 66}, 075004 (2002).



\bibitem{Ellis:1978xg} 
  J.~R.~Ellis, M.~K.~Gaillard and D.~V.~Nanopoulos,
  Phys.\ Lett.\ B {\bf 80}, 360 (1979)
  [Erratum-ibid.\  {\bf 82B}, 464 (1979)]
.

\bibitem{Hisano:1992jj} 
  J.~Hisano, H.~Murayama and T.~Yanagida,
  Nucl.\ Phys.\ B {\bf 402}, 46 (1993)
.

\bibitem{Aoki:2006ib} 
  Y.~Aoki, C.~Dawson, J.~Noaki and A.~Soni,
  Phys.\ Rev.\ D {\bf 75}, 014507 (2007)
.

\bibitem{GUT2012}
 Y.~Aoki, Talk at GUT2012, 16 March, 2012.

\bibitem{Claudson:1981gh} 
  M.~Claudson, M.~B.~Wise and L.~J.~Hall,
  Nucl.\ Phys.\ B {\bf 195}, 297 (1982).


\bibitem{Aoki:2008ku} 
  Y.~Aoki {\it et al.}  [RBC-UKQCD Collaboration],
  Phys.\ Rev.\ D {\bf 78}, 054505 (2008)
.

\bibitem{Siegel:1979wq} 
  W.~Siegel,
  Phys.\ Lett.\ B {\bf 84}, 193 (1979).

\bibitem{Bjorkman:1985mi} 
  J.~E.~Bjorkman and D.~R.~T.~Jones,
  Nucl.\ Phys.\ B {\bf 259}, 533 (1985).

\bibitem{Machacek:1983tz} 
  M.~E.~Machacek and M.~T.~Vaughn,
  Nucl.\ Phys.\ B {\bf 222}, 83 (1983).
\bibitem{Abbott:1980zj} 
  L.~F.~Abbott and M.~B.~Wise,
  Phys.\ Rev.\ D {\bf 22}, 2208 (1980).

\bibitem{Munoz:1986kq} 
  C.~Munoz,
  Phys.\ Lett.\ B {\bf 177}, 55 (1986).


\bibitem{Ghilencea:1997mu} 
  D.~Ghilencea, M.~Lanzagorta and G.~G.~Ross,
  Nucl.\ Phys.\ B {\bf 511}, 3 (1998)
.

\bibitem{Randall:1998uk} 
  L.~Randall and R.~Sundrum,
  Nucl.\ Phys.\ B {\bf 557}, 79 (1999);\\
%
  G.~F.~Giudice, M.~A.~Luty, H.~Murayama and R.~Rattazzi,
  JHEP {\bf 9812}, 027 (1998)
.

\bibitem{ArkaniHamed:2004fb} 
  N.~Arkani-Hamed and S.~Dimopoulos,
  JHEP {\bf 0506}, 073 (2005); \\
  G.~F.~Giudice and A.~Romanino,
  Nucl.\ Phys.\ B {\bf 699}, 65 (2004)
  [Erratum-ibid.\ B {\bf 706}, 65 (2005)];\\
  N.~Arkani-Hamed, S.~Dimopoulos, G.~F.~Giudice and A.~Romanino,
  Nucl.\ Phys.\ B {\bf 709}, 3 (2005).

\bibitem{Gabbiani:1996hi} 
  F.~Gabbiani, E.~Gabrielli, A.~Masiero and L.~Silvestrini,
  Nucl.\ Phys.\ B {\bf 477}, 321 (1996)
.

\bibitem{Hisano:2006nn} 
  J.~Hisano, S.~Matsumoto, M.~Nagai, O.~Saito and M.~Senami,
  Phys.\ Lett.\ B {\bf 646}, 34 (2007)
.

\bibitem{Hisano:2010fy} 
  J.~Hisano, K.~Ishiwata and N.~Nagata,
  Phys.\ Lett.\ B {\bf 690}, 311 (2010); \\
%
  J.~Hisano, K.~Ishiwata, N.~Nagata and T.~Takesako,
  JHEP {\bf 1107}, 005 (2011);\\
  T.~Moroi and K.~Nakayama,
  Phys.\ Lett.\ B {\bf 710}, 159 (2012)
.

\bibitem{Nihei:1994tx} 
  T.~Nihei and J.~Arafune,
  Prog.\ Theor.\ Phys.\  {\bf 93}, 665 (1995)
.


\end{thebibliography}
\end{document}